\documentstyle[prb,aps,multicol,epsf]{revtex}
\renewcommand{\narrowtext}{\begin{multicols}{2}
\global\columnwidth20.5pc}
\renewcommand{\widetext}{\end{multicols}
\global\columnwidth42.5pc} \multicolsep = 8pt plus 4pt minus 3pt

\input{psfig.sty}

\begin{document}

\draft
\title{Spin Dynamics of Cavity Polaritons}

\author{M. D. Mart\'in and L. Vi\~na}
\address{
Departamento de Fisica de Materiales C-IV, Universidad Autonoma de Madrid,
Cantoblanco,
E-28049 Madrid, Spain}

\author{J.K. Son and E.E. Mendez}
\address{
Department of Physics and Astronomy, SUNY at Stony Brook, N.Y. 11794-3800, USA}
\date{\today}
\maketitle
\begin{abstract}
We have studied polariton spin dynamics in a GaAs/AlGaAs microcavity by means of polarization- 
and time-resolved photoluminescence spectroscopy as a function of excitation density and 
normal mode splitting. The experiments reveal a novel behavior of the degree of polarization of the 
emission, namely the existence of a finite delay to reach its maximum value. We have also found that 
the stimulated emission of the lower polariton branch has a strong influence on spin dynamics: in an interval 
of $\sim$150 ps the polarization changes from +100$\%$ to negative values as high as -60$\%$. 
This strong modulation of the polarization and its high speed may open new possibilities for spin-based devices.
\end{abstract}
\pacs{PACS numbers: 71.36.+c, 78.45.+h, 78.47.+p, 71.35.Gg}

\narrowtext
Semiconductor microcavities are one of the most suitable 
structures to study light–matter interaction. In the strong coupling regime, 
first observed in 1992, excitons and photons form mixed 
states, named cavity polaritons.\cite{Weisbuch92}
The signature of this regime is an anticrossing of the exciton and 
cavity modes when they are brought into resonance. Although cavity 
polaritons have been profusely investigated,\cite{Houdre94,Norris93,Houdre95}
there are some aspects of their optical properties which need better 
understanding, in particular those that refer to non-linear processes and the polarization of the emitted light.

In studying non-linear processes, it has been difficult to maintain the 
polaritonic signature because of saturation of the strong coupling regime.\cite{Houdre95,Pau96,Khitrova99}
This is the case in Vertical Cavity Surface Emitting Lasers (VCSEL's), where 
the nonlinear emission originates in the population inversion of a dense electron-hole 
plasma in the weak coupling regime.\cite{Sale95}
The polaritonic character of stimulated emission in both ${III-V}$\cite{Senellart99} and ${II-VI}$\cite{Dang98} 
microcavities reported recently has been questioned by 
calculations that claim that those results can be explained 
within a fermionic quantum theory.\cite{Khitrova99}

The polarization of the light emitted by bare semiconductor quantum wells 
(QWs) has been widely investigated.\cite{Damen91}
In fact, because the polarization is directly linked to the spin 
(i. e. third component of the total angular momentum), its study is part of a 
new field, known as spintronics, that aims to develop spin-based fast optoelectronic devices. 
The mechanisms responsible for the spin relaxation of excitons in QWs and its 
dependence on different parameters such as well width, temperature and 
excitation density have been established.\cite{Damen91,Maialle93}  
In microcavities, due to the mixed photon-exciton character of the 
polaritons and the inefficiency of those mechanisms on the cavity like mode, 
significant changes on the spin dynamics are to be expected. But, in spite of these 
differences, only a few works on the polarization properties of VCSEL's\cite{Ando98,Gahl99}
and microcavities\cite{Tartakowskii99} 
have been reported, the latest one on the cavity-polaritons spin 
properties in the nonlinear regime and under cw excitation.\cite{Tartakowskii99}
Following a preliminary study of spin dynamics in a semiconductor 
microcavity,\cite{Martin99}
we report in this paper on the time evolution of 
the polariton spin in both linear and non-linear regimes. 
Specifically, we show that the emission is highly polarized in the non-linear regime and 
that the polarization dynamics is strongly influenced by exciton-cavity detuning. 

The microcavity studied in this work, grown by molecular beam epitaxy, 
consists of three GaAs quantum well regions 
embedded in a 3$\lambda$/2 Al$_{0.25}$Ga$_{0.75}$As 
Fabry-Perot resonator clad by dielectric mirrors. The top and bottom 
mirrors are distributed Bragg reflectors made of twenty and a half and twenty-four alternating  
AlAs/Al$_{0.35}$Ga$_{0.65}$As $\lambda$/4 layers, 
respectively. The QWs are placed at the antinodes of the 
resonator's standing wave. A slight variation (introduced by design during growth) 
of the cavity's thickness along the radial direction of the wafer allowed to tune 
the cavity's resonance with the transition in the QWs by moving the excitation 
spot across the sample.\cite{Tao98}
Using low temperature cw-photoluminescence (PL) 
measurements we have identified exciton-like (X) and cavity-like (C) modes, 
whose Normal Mode Splitting (NMS) variation was found to be between 3.5 and 7 meV as 
the laser spot was moved across the sample.

We have used time-resolved spectroscopy to study polariton recombination 
and spin dynamics as a function of excitation density and exciton-cavity 
detuning (E$_{C}$-E$_{X}$). The experiments were performed 
under non-resonant excitation above the 
cavity stop-band (1.71 eV) and the PL emitted by the sample 
was analyzed in a conventional up-conversion spectrometer\cite{Shah88}
with a time resolution of $\sim$ 2 ps. 
The sample was mounted on a cold-finger cryostat where the temperature 
was kept at 5 K. For polarization resolved measurements 
two $\lambda$/4 plates were included in the 
optical path of the experiment. Under 
$\sigma ^+$ excitation, the degree of polarization 
of the PL is defined as 
$\wp=\frac{I^+-I^-}{I^++I^-}$,
where $I^{\pm }$ denotes the PL emitted with $\pm$1 helicity. The analysis of this quantity gives direct information 
about the spin relaxation processes, as it is directly related to the difference of +1 and -1 
spin populations.\cite{Maialle93}
In the following, we will refer to this quantity simply as polarization.

Initial time-resolved experiments under weak excitation 
($I_{exc}<$ 19~$W/cm^2$), confirmed the NMS variation across the sample 
extracted from cw measurements.\cite{Tao98}
The study of the time evolution of both peaks revealed that the NMS does 
not influence the dynamics for positive detunings, in agreement 
with Abram {\it et al.}\cite{Abram96} 
The characteristic rise and decay times of the PL were very similar for 
both polariton branches, and amounted to 
$\tau_r^X \sim 100~ps$,
$\tau_d^X \sim 300~ps$,
$\tau_r^C \sim 70~ps$,
and $\tau_d^C \sim~250 ps$,
where $r$ ($d$) denotes rise (decay) time.

With increasing excitation density, drastic changes were observed in the time-resolved spectra as well as in the recombination dynamics. 
At low power densities, both the lower (LPB) and upper (UPB) polariton branches have a similar dependence on power 
(slightly larger that linear). This dependence is maintained for the UPB in the whole range of excitation densities used in our 
experiments (up to 40~$W/cm^2$). 
In contrast, the dependence of the LPB emission on power shows a threshold, 
$I_{th}$, at $\sim$ 20~$W/cm^2$ (Ref. \cite{Tassone97}).

\begin{figure}
\psfig{figure=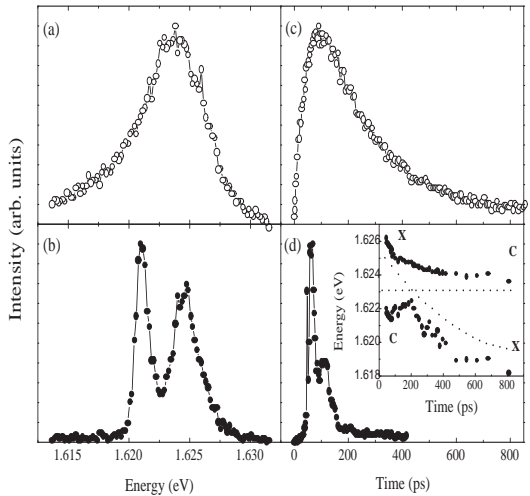,height=8.0cm,width=8cm}
\caption{(a) PL spectra measured 100 ps after excitation with a 0.35 $I_{th}$ density.
(b) PL spectra measured 100 ps after excitation with an $I_{th}$ density.
(c) Cavity-mode time evolution for an excitation density of 0.35 $I_{th}$.
(d) Cavity-mode time evolution for an excitation density of $I_{th}$. 
Inset: energy positions of both polariton branches as a function of time for a density $I_{th}$. 
The lines are a guide to the eye.}
\label{PL}
\end{figure}

Figures 1a and 1b display two PL spectra measured 100 ps after excitation, below ({\Large $\circ$}, 7~$W/cm^2$) 
and at the threshold ({\Large $\bullet$}, 20~$W/cm^2$), respectively. At low power, the spectrum is dominated by the 
UPB at 1.624 eV (Fig. 1a). The LPB becomes resolvable for I $\geq$14~$W/cm^2$ as a very narrow peak at 
1.621 eV (Fig. 1b), and it sharpens with increasing density, up to $I_{th}$, reducing its 
width by a factor of $\sim 4$. 
This narrowing also occurs for the UPB, but its linewidth is only reduced by a factor for $\sim 2$. 
The NMS is practically independent of power.

The time evolution is also strongly affected by an increase of excitation density, 
as shown in Figures 1c and 1d. For small excitation densities 
({\Large $\circ$}, 7~$W/cm^2$) the time evolution is similar to that typical of QWs under non-resonant excitation: 
the emission is characterized by slow rise and decay times.\cite{Damen91} 
For larger excitation ({\Large $\bullet$}, 20~$W/cm^2$), the rise and decay times are faster and there is a 30 ps delay 
before any PL from the sample is observable. This onset in the PL is due to the bottleneck in the 
relaxation of polaritons towards \textbf{K}$_{\parallel}$ = 0 states. \cite{Tassone97}

The subject of non-linear emission in semiconductor microcavities has been controversial regarding the 
existence of polaritons, due to bleaching 
at high densities.\cite{Pau96,Khitrova99,Senellart99,Tartakowskii99} 
In our experiments, excitons and photons seem to be still strongly coupled above 
$I_{th}$, as can be inferred from the anticrossing depicted in the inset of Fig. 1d. 
The non-resonantly created excitons relax their energy very rapidly 
towards \textbf{K}$_{\parallel}$ {$\simeq$} 0 states and, a few picoseconds 
after excitation, the X mode is observed at 1.626 eV. At these very short times, the LPB (1.622 eV) is photon-like. 
With increasing delay, the X mode red shifts due to the decrease of polariton density, similarly to the behavior of excitons in 
bare QWs. \cite{Schmitt-Rink85}
However, since the C mode energy is density independent, both modes become resonant, and a clear anticrossing is 
observed at $\sim$ 180 ps. At longer times, the LPB (UPB) recovers the X-like (C-like) character determined by cw measurements.

The three effects discussed above, linewidth reduction, excitation density threshold and, especially, the anticrossing in 
time suggest that the PL observed above $I_{th}$ can be attributed to polariton stimulated emission.

Let us now concentrate on polariton spin dynamics. 
A $\sigma ^+$ excitation pulse will initially populate the +1 spin level but a -1 spin population will 
appear as a result of spin flip mechanisms, which eventually balance both spin populations\cite{Maialle93} 
and therefore reduce the polarization to $\wp$=0. For excitons in bare QWs, the polarization reaches its 
maximum value just after excitation and then decays exponentially to zero.\cite{Damen91} 
On the other hand, in microcavities, due to the complex nature of polaritons, one expects the spin 
dynamics of this mixed state to be different from that of bare excitons or photons. 
This fact is documented in Figure 2, which depicts the time evolution of the polarization of the cavity 
mode for two different excitation densities below the nonlinear emission threshold. 
In contrast with the monotonically decreasing behavior of $\wp$ found in 
bare QWs, in our microcavity a 
maximum is observed at a finite time after excitation. 
The polarization at t = 3 ps is $\sim$ 10$\%$, which means that, after the relaxation of 
polaritons to \textbf{K}$_{\parallel}$ {$\simeq$} 0 
states, only 55$\%$ of the total population is in the +1 spin state. Such a small value of 
$\wp$ is mainly due to the non-resonant excitation conditions. $\wp_{max}$ is reached in 60-100 ps, and its 
value increases with excitation density, being as high as 
80$\%$ (Fig. 2(b), 19~$W/cm^2$) before entering the stimulated emission regime. 
These findings corroborate recent cw-results that show that the polariton system 
can be markedly spin polarized.\cite{Tartakowskii99}

\begin{figure}
\psfig{figure=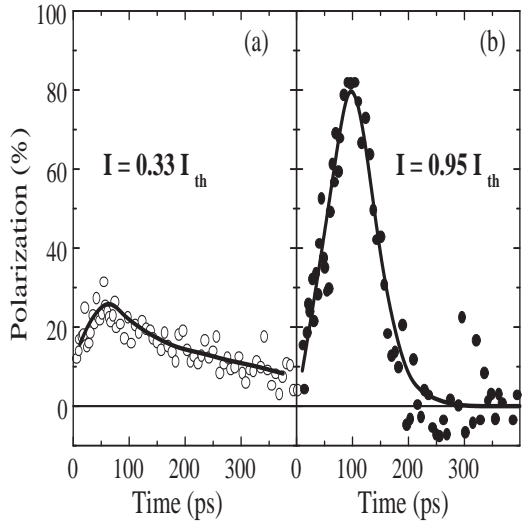,height=8.5cm,width=8cm}
\caption{Time evolution of the cavity-mode polarization for excitation densities: 
0.33 $I_{th}$ (a) and 0.95 $I_{th}$ (b).}
\label{polarization}
\end{figure}

The fact that a finite time is needed to reach $\wp_{max}$ 
implies that there must be a new scattering mechanism that favors polaritons with +1 spin, 
and thus competes with spin relaxation and  tends to prevent equalization of both spin populations. 
The relaxation of large in-plane wave vector excitons, via emission of acoustic phonons, 
is stimulated by the polariton final-state population. \cite{Baars00} 
The increase of $\wp_{max}$  with excitation density (25$\%$ @ 0.33 $I_{th}$, 80$\%$ @ 0.95 $I_{th}$) 
evidences that there is an enhancement of the scattering to the +1 spin state. 
The stimulation does not occur for the $\sigma ^-$ polarized LPB emission, 
which also shows a time evolution with rise and decay times much longer than those observed 
in the non-linear regime for the $\sigma ^+$ emission. 
This process is not only spin selective but also induces an increase of the +1 population by flipping 
the spin of minoritary polaritons (-1).

An additional fact evidences the importance of the "polaritonic" stimulation on the spin behavior: 
the decrease in the time needed to reach 
$\wp_{max}$ with increasing positive detuning. This means 
that as we move away from resonance and the excitonic 
component of the LPB increases, the time evolution of  $\wp$ approaches that characteristic of bare excitons, 
with the maximum value of $\wp$ occurring closer to t=0. 
The explanation given here is only qualitative and a complete theoretical description must be developed before this 
new mechanism of scattering into spin polarized states can be fully understood.

Recently, a rate-equation model has been successfully applied to describe the optical properties of 
cavity-polaritons in the nonlinear regime, under cw excitation, \cite{Pau96} 
and a microscopic fermionic many-body theory has been developed to explain the linear and non-linear 
behavior of normal-mode coupling in microcavities, including dynamics of the light emission.\cite{Khitrova99}
However, even in these models, the spin of the polaritons was not taken into account. 
Our results provide new valuable information on the stimulated scattering into spin-polarized states of the LPB, 
and should attract interest for theoretical work that includes the spin in the calculations.

For excitation densities above the threshold, the time evolution of the cavity-mode polarization displays a 
behavior even more surprising. The LPB polarization reaches values as high as 
95$\%$ when entering into the nonlinear emission regime. In contrast, for the UPB, although it shows a 
similar behavior, its polarization is only 60$\%$. After the initial rise of the polarization, once the maximum is 
reached, its dynamics is strongly dependent on NMS. Figure 3 depicts the time evolution 
of $\wp$ for two different points of the sample, with different NMS, under an excitation density of 
2 $I_{th}$. For small exciton-cavity detunings (Fig. 3(a), 4.5 meV) a negative dip (-60$\%$) 
is observed at $\sim$150 ps, which is absent for larger NMS (Fig. 3(b), 6 meV).

\begin{figure}
\psfig{figure=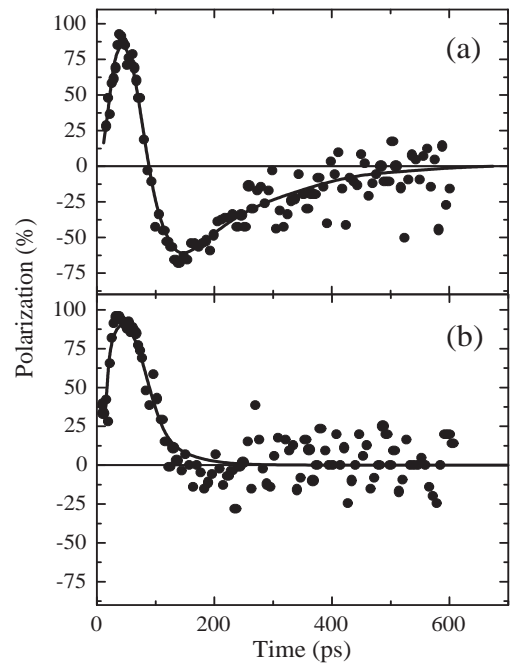,height=10.0cm,width=8cm}
\caption{Time evolution of the cavity-mode polarization for excitation 
densities $\sim$2 $I_{th}$ measured at different points of the sample, 
characterized by a NMS of (a) 4.5 meV and (b) 6 meV.}
\label{normal_mode}
\end{figure}

The negative polarization is a consequence of the fast disappearance of the +1 polaritons, due to the 
stimulated $\sigma ^+$ emission of the LPB and the concurrent slower dynamics of the 
$\sigma ^-$ emission. The -1 polariton population overcomes that of +1 spin due to the lack of stimulation for 
$\sigma ^-$ polarization and also because the spin-flip processes are not fast enough to compensate the emptying of the +1 polaritons. 
The remarkable change in the state of polarization of the emitted light from +80$\%$ to -60$\%$, taking place in a very short time 
($\sim$100 ps), is unique and, to the best of our knowledge, has not been reported before in any semiconductor based system.

Once the minimum of $\wp$ is reached, the polarization dynamics becomes slower: by then the polariton population has decreased by a 
factor 5 to 10 (depending on power density) and the remaining +1 spin population is too small to give rise to stimulation. 
Under these conditions, only the usual spin-flip mechanisms govern the polarization, which decreases steadily. Figure 3b shows that the 
negative dip has disappeared for larger NMS, due to the modification of the stimulated emission dynamics. 
The decay time of the $\sigma ^+$ PL becomes slower and the loss of +1 polaritons is neutralized by 
flipping -1 spins and as a result $\wp$ does not reach negative values. One can also observe in Fig. 3b that the 
abrupt decay of the polarization is slowed down with increasing NMS.

It should be mentioned that excitation with $\sigma ^-$ yields identical results to those of the 
$\sigma ^+$ excitation discussed above, as expected from time reversal symmetry arguments. 
The sign reversal of the polarization is also observed for the $\sigma ^-$ excitation and it is also the majority spin 
population (-1 in this case) the only one that undergoes stimulation.

In summary, our experiments on polariton recombination as a function of excitation density and exciton-cavity 
detuning have revealed strong nonlinearities in the emission of the lower polariton branch. 
A careful study of the time evolution of the polarization has shown the existence of a new scattering mechanism for the polaritons 
that is spin selective and gives rise to very high values of the polarization. 
The Normal Mode Splitting plays a key role on the spin relaxation of cavity polaritons, leading to a reversal in the 
polarization for small detunings. The large contrast in the polarization and its high speed open the possibility of new 
concepts for spintronic devices, such as ultrafast switches, based on the spin dynamics of microcavity polaritons.

\acknowledgments
We are thankful to Dr. I. W. Tao and R. Ruf  for growing the 
samples used in this work, which has been supported by
Fundaci\'on Ram\'on Areces, the Spanish DGICYT under contract PB96-0085, 
the CAM (07N/0026/1998), the Spain-US Joint Commission and the U.S. Army Research Office.

\vspace{-0.5cm}

\widetext
\end{document}